\documentclass[a4paper]{jpconf}
\usepackage{graphicx}

\begin{document}
\title{The collisional relaxation of electrons in hot flaring plasma and
inferring the properties of solar flare accelerated electrons from X-ray observations}

\author{N. L. S. Jeffrey$^{1}$, E. P. Kontar$^{1}$, A. G. Emslie$^{2}$, and N. H. Bian$^{1}$}

\address{$^1$ SUPA School of Physics \& Astronomy, University of Glasgow, Glasgow, G12 8QQ, UK}
\address{$^2$ Department of Physics \& Astronomy, Western Kentucky University, Bowling Green, KY, 42101, USA}

\ead{natasha.jeffrey@glasgow.ac.uk}

\begin{abstract}
X-ray observations are a direct diagnostic of fast electrons produced in solar flares, energized during the energy release process and directed towards the Sun. Since the properties of accelerated electrons can be substantially changed during their transport and interaction with the background plasma, a model must ultimately be applied to X-ray observations in order to understand the mechanism responsible for their acceleration. A cold thick target model is ubiquitously used for this task, since it provides a simple analytic relationship between the accelerated electron spectrum and the emitting electron spectrum in the X-ray source, with the latter quantity readily obtained from X-ray observations. However, such a model is inappropriate for the majority of solar flares in which the electrons propagate in a hot megaKelvin plasma, because it does not take into account the physics of thermalization of fast electrons. The use of a more realistic model, properly accounting for the properties of the background plasma, and the collisional diffusion and thermalization of electrons, can alleviate or even remove many of the traditional problems associated with the cold thick target model and the deduction of the accelerated electron spectrum from X-ray spectroscopy, such as the number problem and the need to impose an {\it ad hoc} low energy cut-off.
\end{abstract}

\section{Introduction and the observation of solar flare X-rays}\label{intro}
\begin{figure}
\centering
\includegraphics[width=8.5 cm]{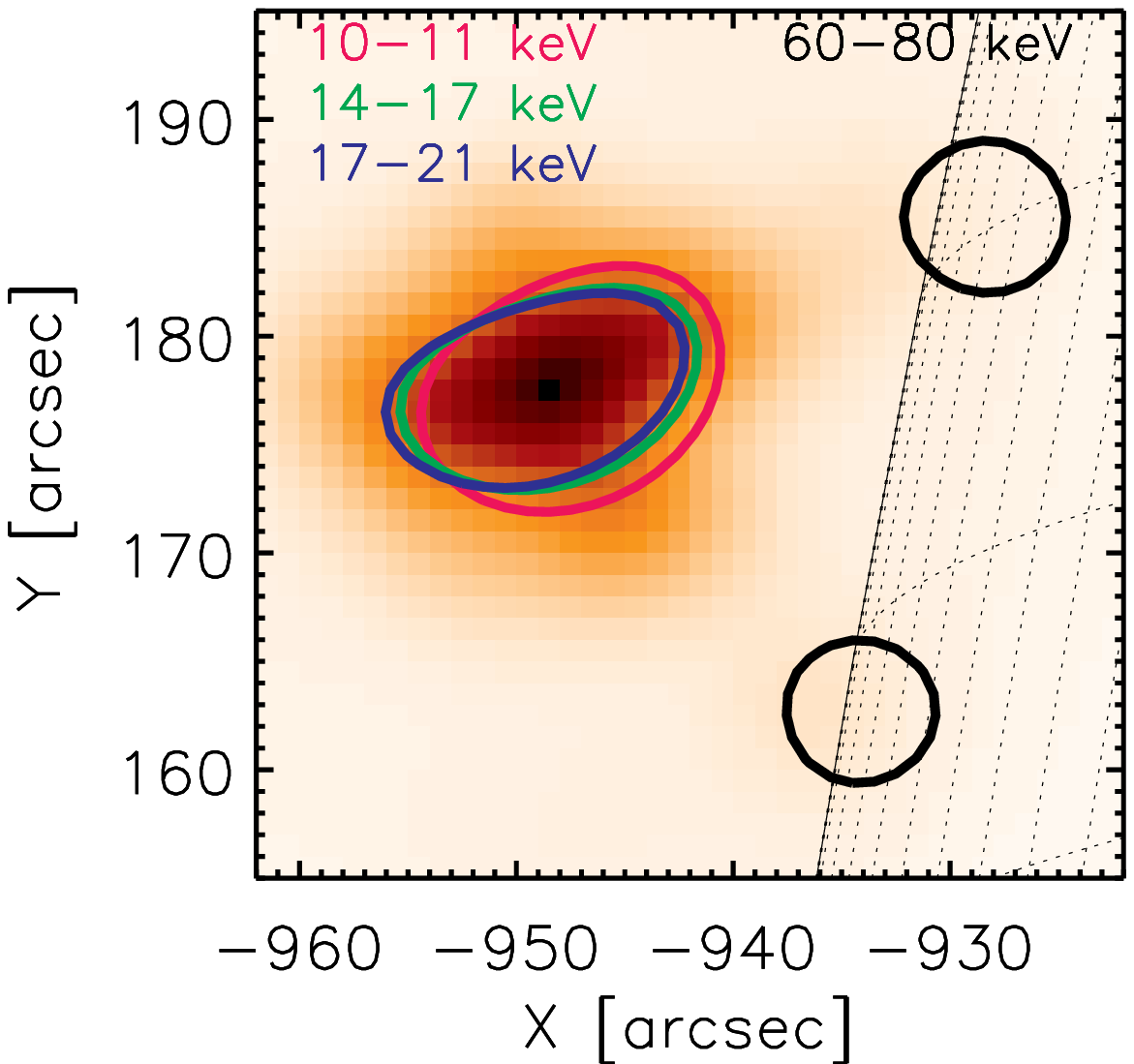}\hspace{-50pt}
\includegraphics[width=8.5 cm]{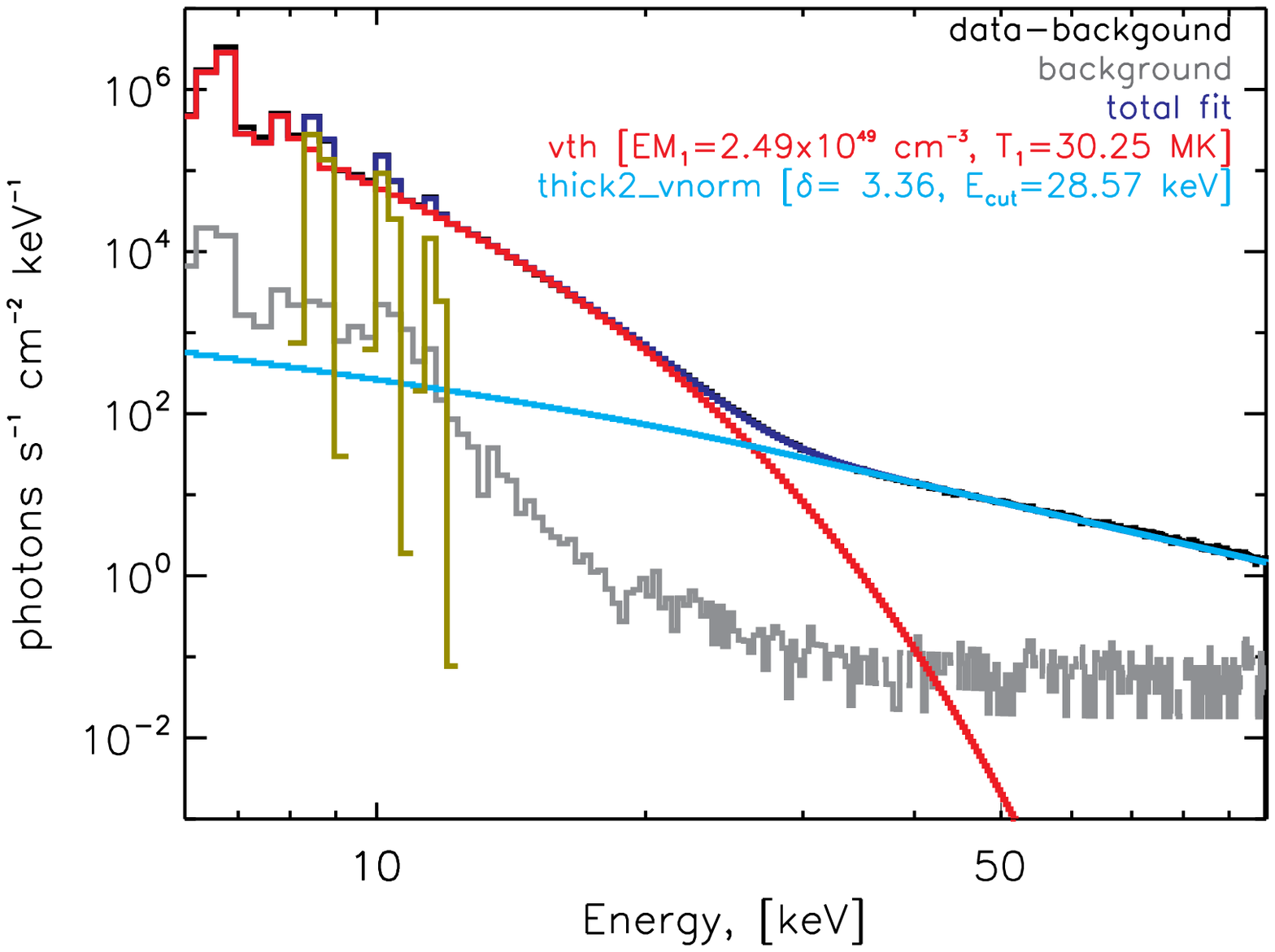}\vspace{-20pt}
\includegraphics[width=8.5 cm]{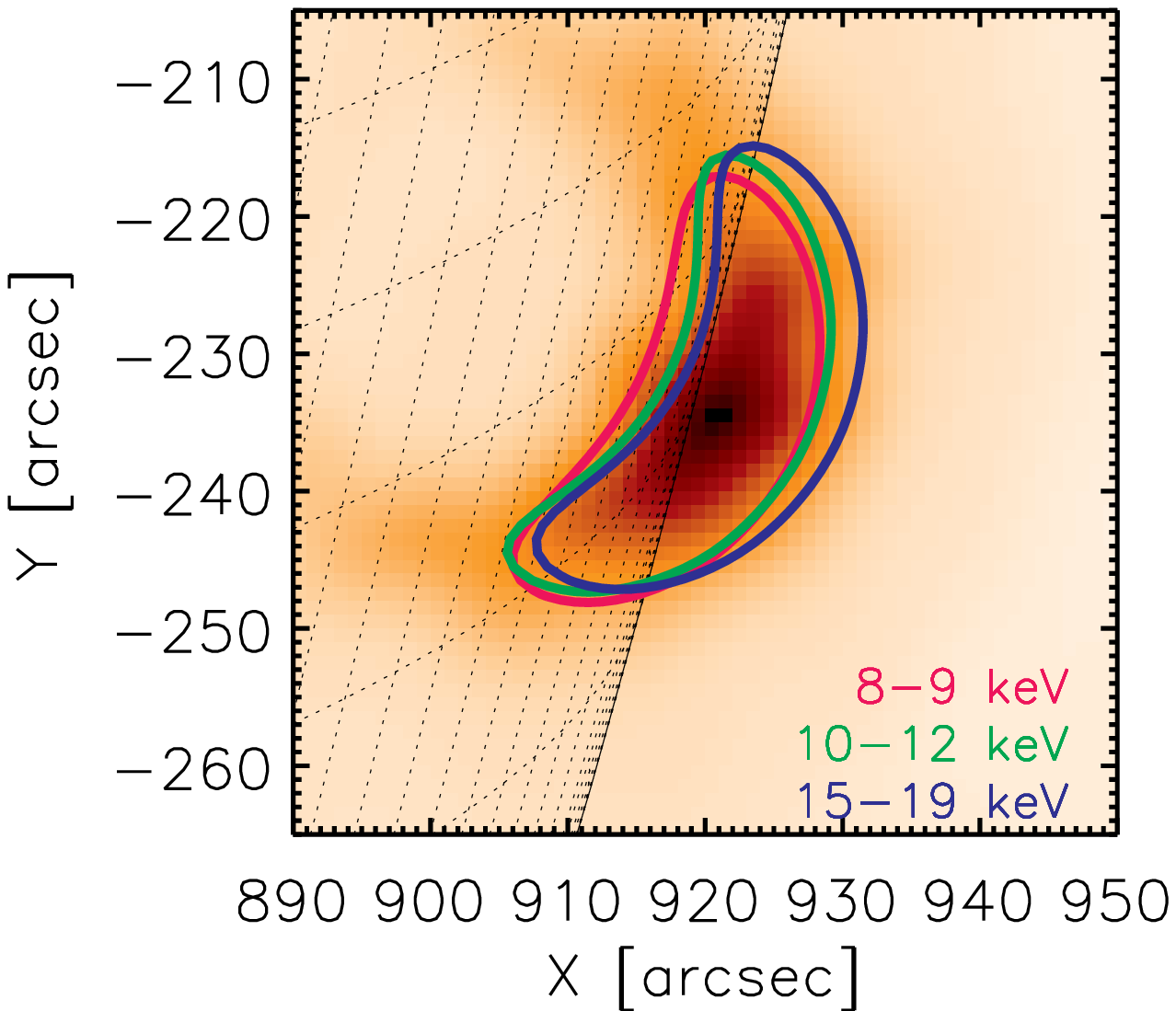}\hspace{-50pt}
\includegraphics[width=8.5 cm]{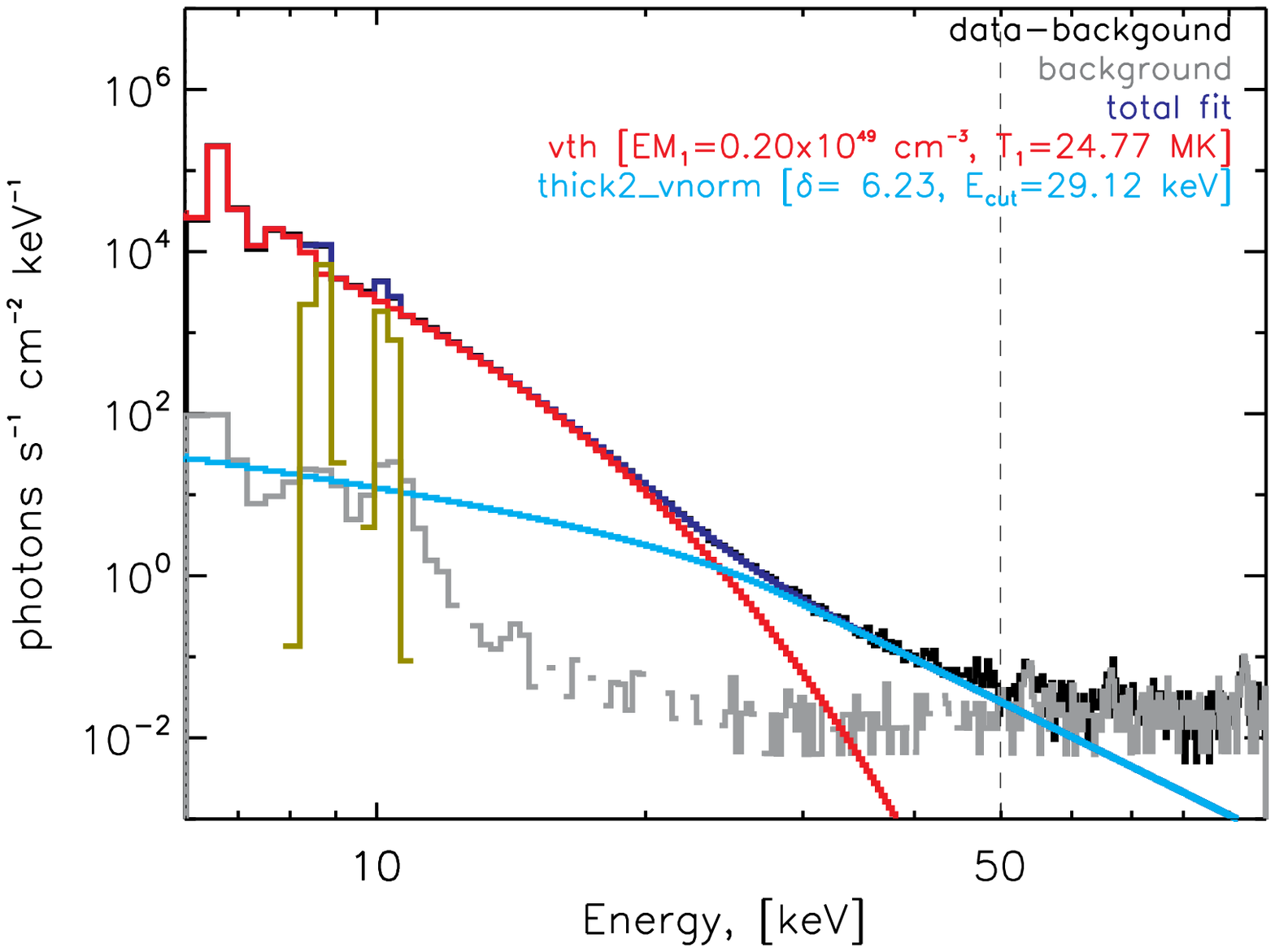}
\caption{{\it Left:} {\em RHESSI\,} X-ray images of two solar flares, where the top image shows a `standard' flare type (SOL2013-05-13T16:04) with a hot coronal source observed at energies of $\sim10$ to $25$ keV and X-ray footpoints above $\sim30$ keV located lower in the atmosphere. The bottom image shows a coronal dominated flare type (SOL2005-08-23T14:30), where both the thermal and nonthermal emission exist cospatially. {\it Right:} {\em RHESSI\,} X-ray spectra for each flare. Most flares, independent of the flare type have two main components: a hot thermal component dominant at lower energies (red) and a power law nonthermal component at higher energies (blue).}
\label{fig1}
\end{figure}

During a solar flare, a very large number of electrons, of the order of $10^{36}$ electrons~s$^{-1}$ are accelerated, but the mechanism responsible for their acceleration remains poorly understood (see \cite{2011SSRv..159..357Z} for a recent review). The properties of accelerated electrons are mainly deduced by X-ray observations, currently using X-ray imaging and spectroscopy provided by the {\em Ramaty High Energy Solar Spectroscopic Imager (RHESSI)} \cite{2002SoPh..210....3L}. Solar flare X-rays are mainly produced as bremsstrahlung from electron-ion collisions and are optically thin in the solar atmosphere where hydrogen number densities range from around $10^{8}-10^{16}$ cm$^{-3}$. X-ray observations thus provide an untainted inference of the {\it emitting} electron distribution, which is generally different from the {\it accelerated} electron distribution, since the latter can be modified en route by many processes in the solar atmosphere. These processes include Coulomb collisions (mainly with other electrons), and also non-collisional and turbulent processes (for example \cite{2014ApJ...780..176K,2012A&A...539A..43K}). Apart from rare cases, such as the {\it `cold flare'} \cite{2011ApJ...731L..19F}, most solar flare spectra consist of two parts, usually treated as separate entities: a bremsstrahlung power law component dominant above $\sim$25 keV that is produced by the collision and deceleration of fast accelerated electrons, and a thermal component (mostly consisting of bremsstrahlung, but also free-bound and line emission), produced by a highly heated plasma (1 keV or more) that completely dominates the X-ray spectrum at energies below $\sim$25 keV (see Figure \ref{fig1}) \cite{2011SSRv..159..301K,2011SSRv..159..107H}. {\em RHESSI\,} X-ray imaging usually shows that the bulk of hot material $\sim20$ MK ($\sim1.7$ keV) emanates from a coronal source and the nonthermal X-rays from `footpoint' sources, located lower in the atmosphere (see Figure \ref{fig1}). There is also some evidence of hot emission of around 10 MK ($\sim0.86$ keV) closer to the footpoints \cite{2013ApJ...767...83G}. Sometimes the thermal and nonthermal components are cospatial and are both produced in the hot, dense coronal loop, e.g. \cite{2008ApJ...673..576X,2011ApJ...730L..22K,2013ApJ...766...28G} (see Figure \ref{fig1}).

\section{Finding the flare accelerated electron distribution}\label{cold}
\begin{figure}
\centering
\includegraphics[width=7 cm]{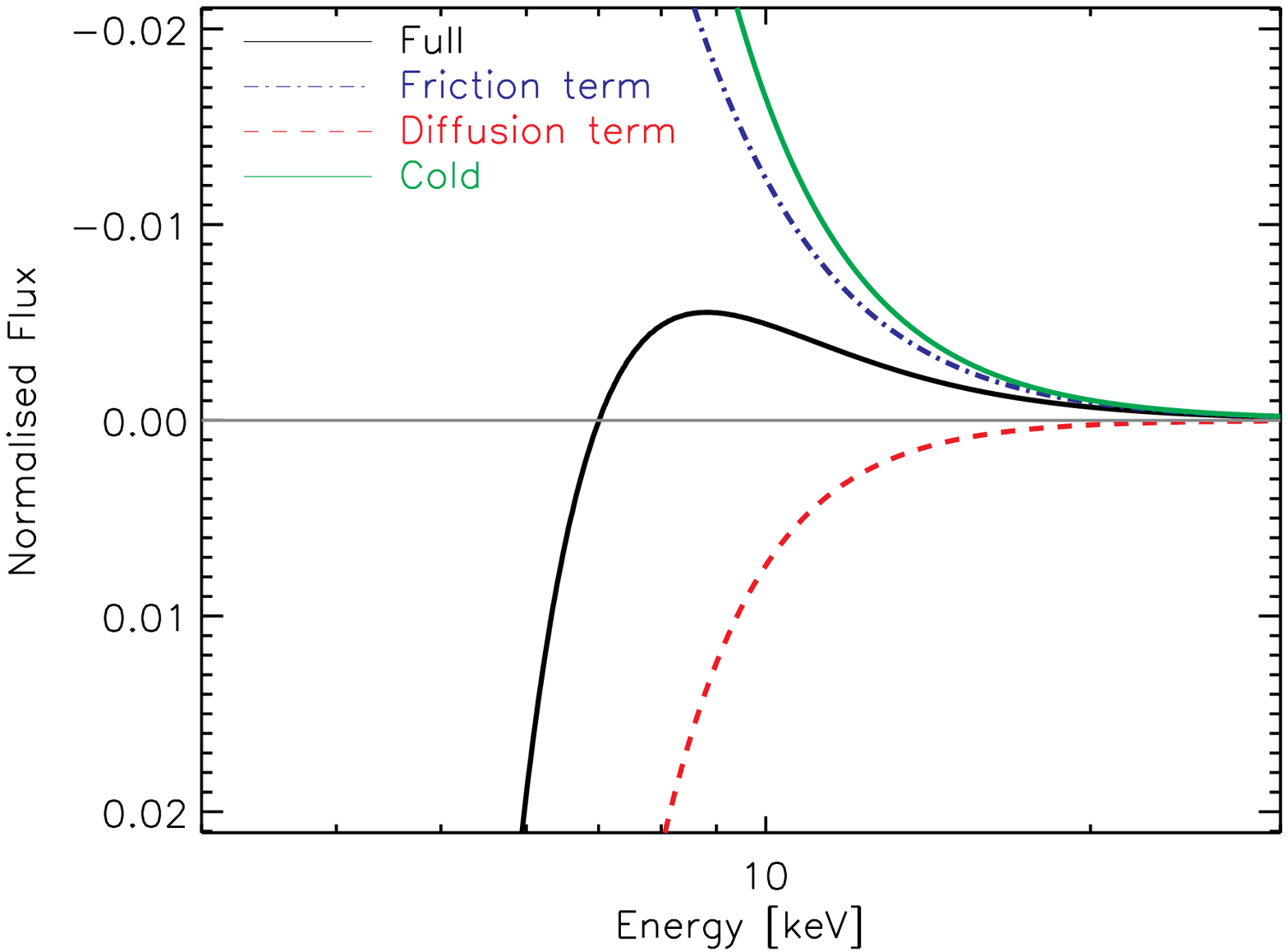}
\includegraphics[width=7 cm]{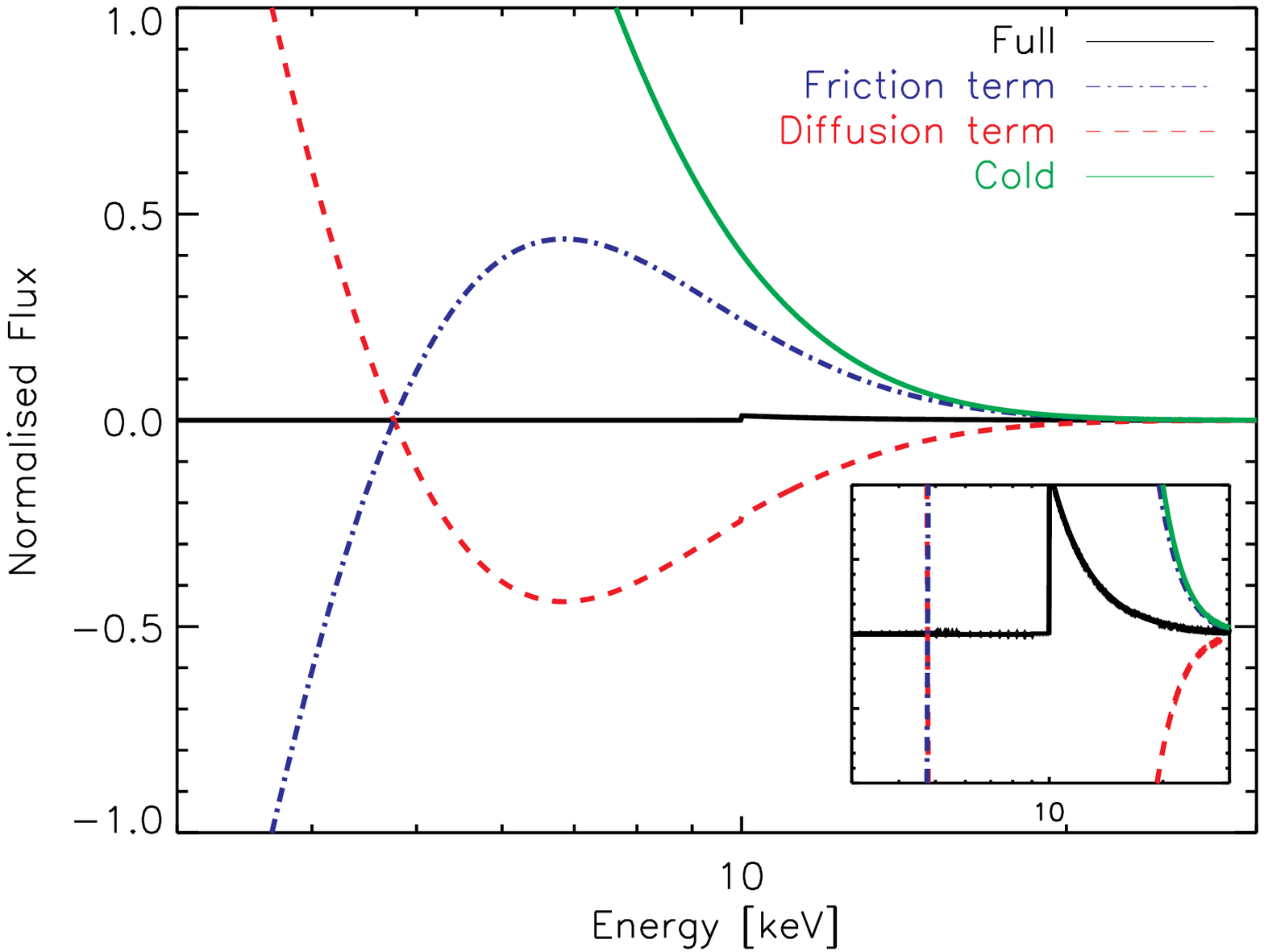}
\caption{Equations (\ref{cttm}) and (\ref{full}) plotted for {\it left:} a power law $\langle nVF\rangle  (E)$, and {\it right:} a Maxwellian plus power law $\langle  nVF\rangle  (E)$, where {\it blue:} frictional first order term (Equation~(\ref{full})), {\it red:} diffusive second order term (Equation~(\ref{full})) and {\it black:} both terms (Equation~(\ref{full})). The green curve represents the cold model (Equation~(\ref{cttm})). In the full model, the diffusive term at low energies cancels the frictional term providing a natural cutoff to the accelerated electron distribution, removing the low energy cutoff problem.}
\label{fig2}
\end{figure}

Commonly, either by a forward fitting or inversion technique, the cold thick target model e.g. \cite{1971SoPh...18..489B,1972SvA....16..273S,1978ApJ...224..241E} is applied to {\em RHESSI\,} X-ray spectra. This model is popular due to its simple analytical form. It assumes that: {\it 1. the accelerated electrons are transported through a `cold' material, whereby the accelerated electron energy $E\gg T$, where $T$ is the temperature of the background material in keV,}  and {\it 2. the observed X-ray bremsstrahlung emission is produced in a region of sufficiently high density that the accelerated electrons lose all of their energy collisionally by friction in this region.} In this model, the dynamics of X-ray emitting electrons is dominated by deterministic collisional energy loss and, hence, the spatially averaged electron spectrum
$\langle nVF\rangle$ \cite{2003ApJ...595L.115B} is related to the accelerated electron spectrum $F_0$ injected into the source via,
\begin{equation}\label{cttm}
F_0(E_0) = - \frac{K}{A}{d \over dE} \biggl [\frac{\langle nVF \rangle (E)}{E} \biggr ]_{E=E_0},
\end{equation}
where $E$ and $E_0$ are the observed and injected electron energies respectively, $K$ is the collisional parameter and $A$ is the injection area. Equation~(\ref{cttm}) is plotted in Figure \ref{fig2} for two different $\langle nVF\rangle $. Solar flare X-ray spectra, in the nonthermal domain, are typically quite steep (Figure \ref{fig1}). Using the above model that only retains the effect of deterministic energy loss requires that
the injected electron flux spectra $F_0(E_0)$ are similarly steep. Since the total injected power is a diverging quantity for such steep power-laws, the concept of a `low-energy cutoff' is frequently used to keep the total number and power of the accelerated electron distribution finite. Such a quantity is also difficult to determine from observation due to the large thermal component dominant at lower energies (Figure \ref{fig1}).

\begin{figure}
\centering
\vspace{-60pt}
\includegraphics[width=14 cm]{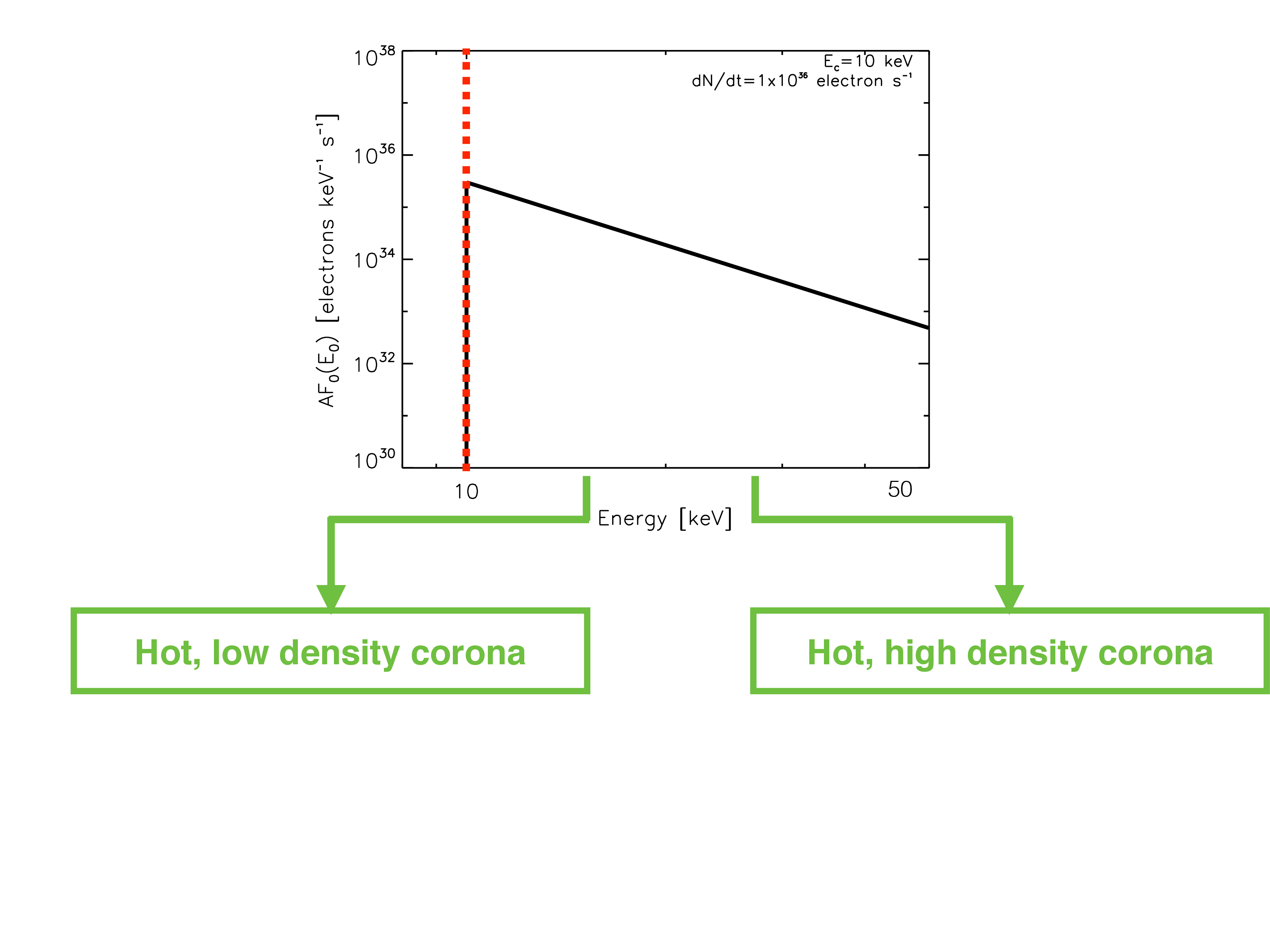}\vspace{-85pt}\hspace{60pt}
\includegraphics[width=15 cm]{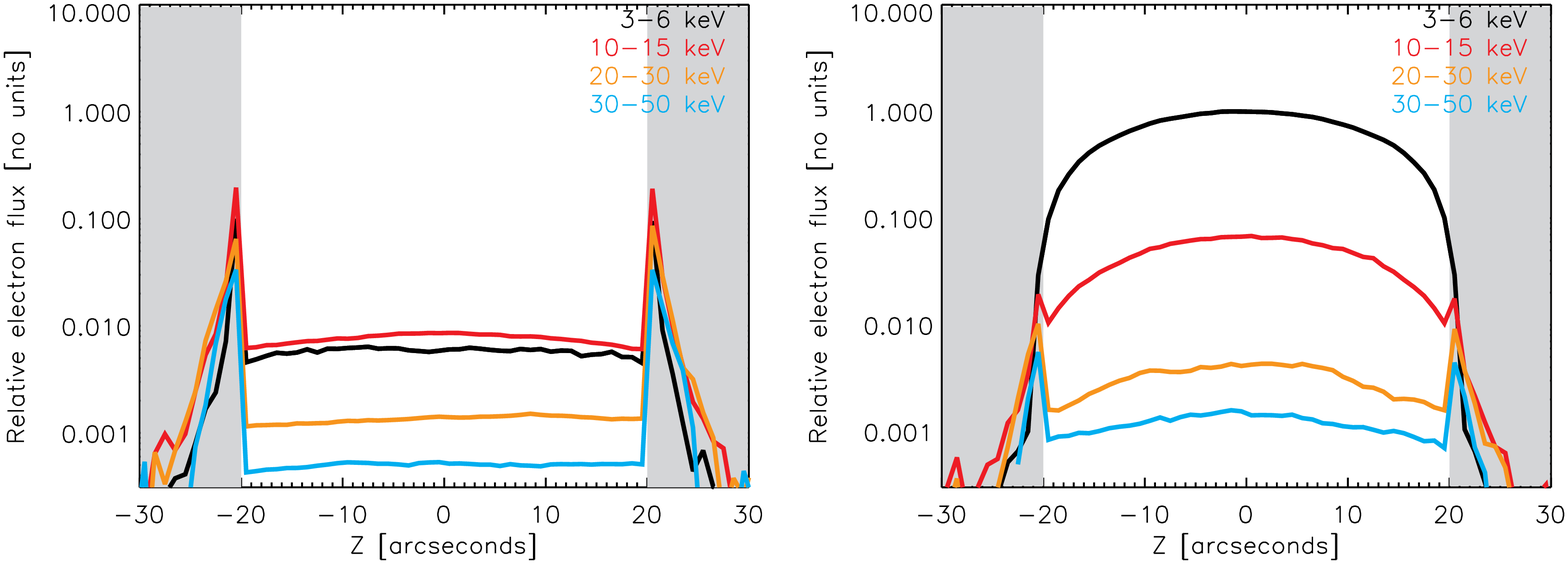}\hspace{60pt}
\includegraphics[width=7. cm]{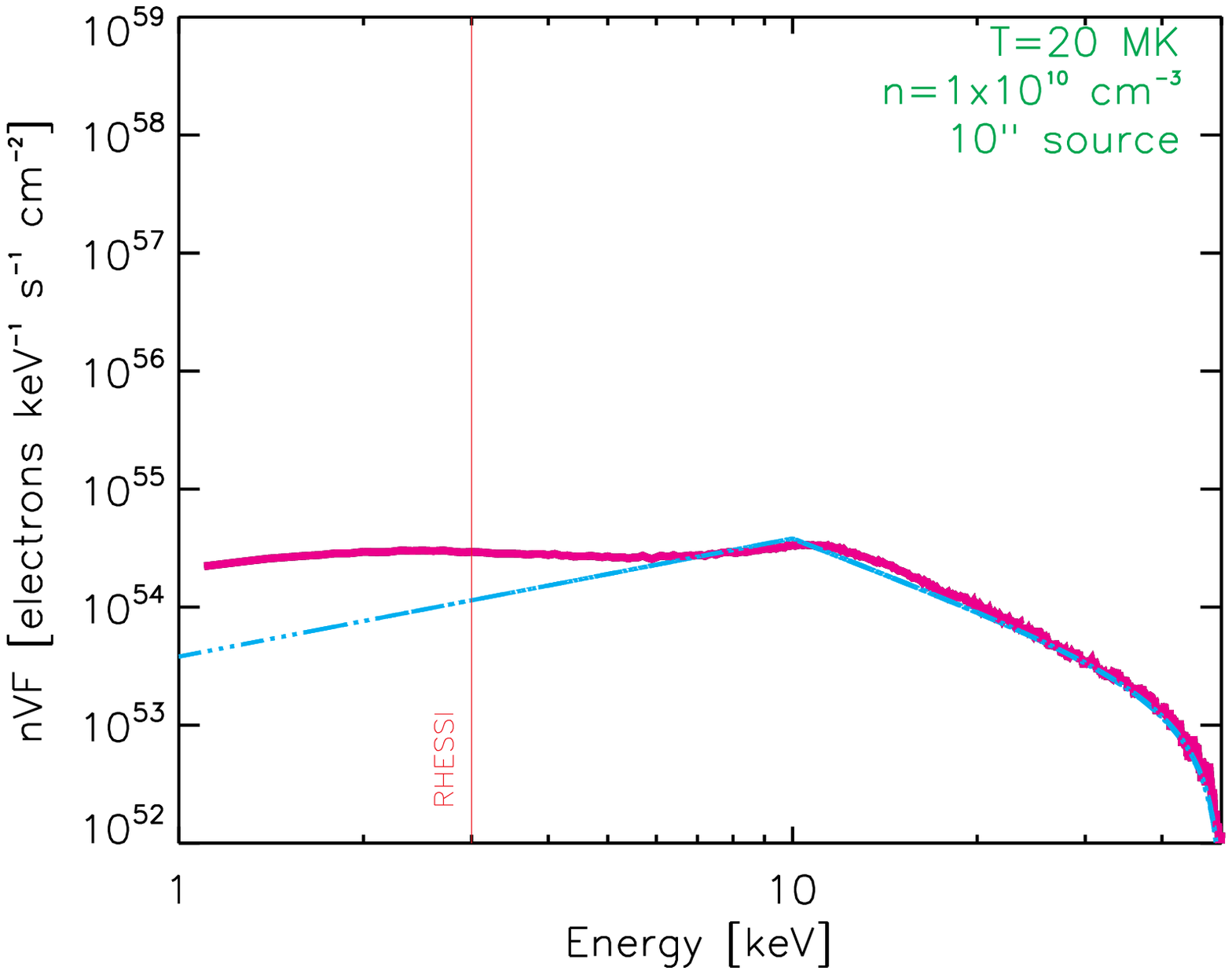}\hspace{10pt}
\includegraphics[width=7. cm]{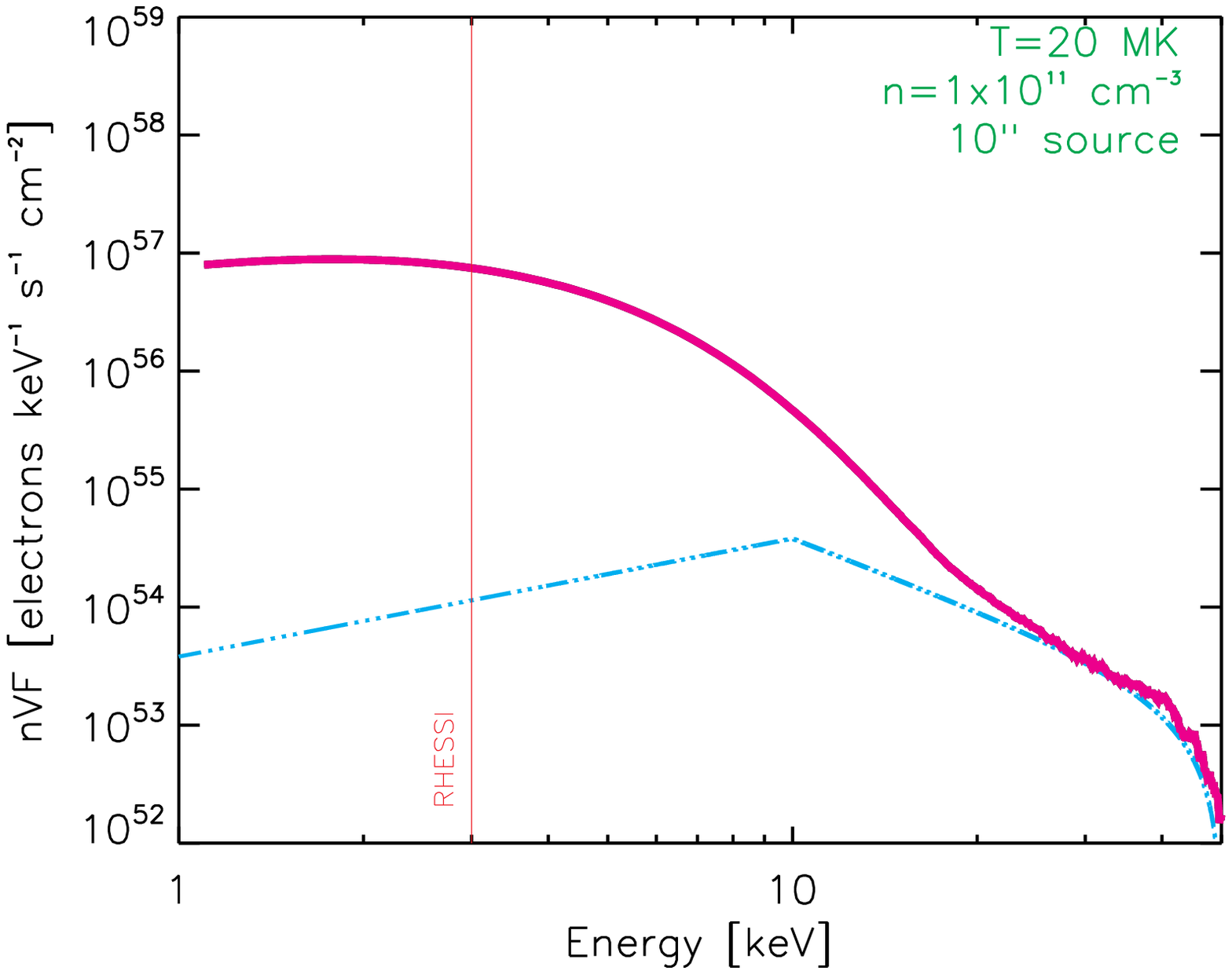}
\caption{The resulting spatial distributions (parallel to a guiding magnetic field) and spatially integrated spectra for a simulated accelerated electron distribution (top panel) transported through two different solar flare coronal environments (middle panels) of {\it left}: $T = 20$ MK and $n = 10^{10}$ cm$^{-3}$, and {\it right}: $T = 20$ MK and $n =10^{11}$ cm$^{-3}$, and injected across $10''$ at the loop top apex ($z=0''$). The grey regions denote a low temperature, high density `chromospheric' region. High coronal densities lead to the bulk of electrons stopping in the corona (coronal thick-target flares), but a low density leads to the bulk of electrons stopping in the dense low atmosphere (footpoint dominated flares). {\it Bottom panels:} the resulting X-ray spectrum (pink) is compared with that of the cold thick target model (blue). The form of the spatial distribution and the integrated X-ray spectrum is dependent on the properties of the background plasma and cannot be ignored if a more correct form for the accelerated electron distribution is to be determined from observation.}
\label{fig3}
\end{figure}

\subsection{Hot plasma in the corona}\label{cold_2}
The cold thick target model is not directly applicable to the analysis of hot flaring plasma, since it does not account for the different behaviour and thermalization of electrons at energies close to $E\sim T$. Such electrons are dominated by collisional diffusion and not by systematic energy loss, thermalizing with the background plasma.  In order to improve the estimation of the accelerated electron properties, we must take into account the presence of a finite target temperature and the resultant collisional diffusion of electrons. In such a model, the observed electron spectrum $\langle nVF\rangle (E)$ and the accelerated electron spectrum $F_0$ are related by

\begin{equation}\label{full}
F_0(E_{0}) = - \frac{2K} { A} \, \frac{d}{dE} \left [ G \left (\sqrt{\frac{E}{T}} \, \right ) \, \left \{ \frac{d \langle nVF \rangle (E)}{dE}
+\frac{1}{E} \, \left ( \frac{E}{T} - 1 \right ) \, \langle nVF \rangle (E) \right \} \right ]_{E=E_0} \,\,\,,
\end{equation}
where $u=\sqrt{E/T}$, $G$ is known as the Chandrasekhar function (see \cite{2014ApJ...787...86J} for more details) and temperature $T$ is in units of keV. Equation~(\ref{full}) reduces to the cold target form, in the high energy, low temperature limit when $E\gg T$, and has two very important distinctions compared to the cold target equation (Equation~(\ref{cttm})). First, the first order friction term is modified so that electrons can either lose or gain energy during a collision, which is particularly important at $E\le T$ (see \cite{2003ApJ...595L.119E}), and {\it even more importantly}, there is the inclusion of a second order term that describes energy changes by collisional diffusion, controlling the behaviour of electrons at $E \sim T$, and ultimately leading to thermalization. Equations~(\ref{cttm}) and (\ref{full}) are shown in Figure \ref{fig2} for two different $\langle nVF\rangle (E)$: a power law and a Maxwellian plus power law. Importantly, Figure~\ref{fig2} shows that the diffusive term cancels the deterministic term at low $E$. For the case of a power law $\langle nVF\rangle (E)$ extending over all $E$, the injected electron flux $F_0(E_0)$ becomes unphysically negative at low injected energies $E_0$, producing a natural cutoff to the injected distribution. In the more plausible `Maxwellian plus power law' case (as might be expected in a hot target), Equation~(\ref{full}) disappears for the Maxwellian component, once again removing the need to apply an arbitrary {\it low energy cutoff} to the accelerated electron distribution.

\subsection{Changes to the observed electron spectrum due to the presence of hot flaring plasma}
In order to show how the presence of hot plasma in a flaring solar atmosphere changes the form of the accelerated electron distribution, Figure~\ref{fig3} plots the resulting spatial distributions and spatially averaged spectra for the same injected electron distribution, for the same target temperature $T=1.7$ keV (20 MK) but with number densities of a) $n=10^{10}$~cm$^{-3}$ and b) $n=10^{11}$~cm$^{-3}$. Unlike in the cold thick target model, the thermalization of fast electrons produces a thermal component at energies $<$10 keV that increases with the background number density $n$. Increasing the background temperature, for a given density $n$, will cause the thermal component to be more pronounced at higher energies (see \cite{2014ApJ...787...86J}). Figure~\ref{fig3} exemplifies how the background properties change the `collisionaly relaxed' electron distribution in the target. From Figure~\ref{fig3}, it is easy to recognize the thermalized part in $\langle nVF\rangle (E)$ at low energies. Importantly, we find that the more realistic model with diffusion correctly describes the electron energy loss rate, particularly at $E \sim T$. Hence, fewer electrons are required to produce a given, observed X-ray flux, of an order of a magnitude (see \cite{2015arXiv150503733K}). The number of electrons required to produce a given X-ray flux falls with increasing temperature $T$ and number density $n$.

\section{Discussion and ongoing work}
The work summarized demonstrates that a more realistic model that includes the effects of collisional diffusion and thermalization of fast electrons, is preferable for inferring the properties of solar flare accelerated electrons at low energies of 10 to 30 keV from X-ray observations. Accounting for such effects can remove problems such as the low energy cutoff, and reduce the number of electrons required to produce a given observed X-ray flux. Importantly, such a model can be applied to all flare types showing the presence of hot plasma, whether they are `standard'  footpoint-dominated or coronal-source-dominated flares. A full description of the work is provided in \cite{2015arXiv150503733K}, and also in \cite{2014ApJ...787...86J}. Further, the application of such a model to real X-ray flare data is currently the topic of ongoing work. We hope it will eventually replace the use of the cold thick target model to deduce the accelerated electron distribution from X-ray spectroscopy.

\subsection{Acknowledgments}
NLSJ was funded by STFC and a SUPA scholarship. EPK and NHB gratefully acknowledge the financial support by an STFC Grant. AGE was supported by NASA Grant NNX14AK56G. NLSJ wishes to thank the organizers for providing her with the opportunity to discuss this work at the conference.

\section{References}

\bibliographystyle{iopart-num}
\bibliography{tampa_references}

\providecommand{\newblock}{}
\begin{thebibliography}{10}
\expandafter\ifx\csname url\endcsname\relax
  \def\url#1{{\tt #1}}\fi
\expandafter\ifx\csname urlprefix\endcsname\relax\def\urlprefix{URL }\fi
\providecommand{\eprint}[2][]{\url{#2}}

\bibitem{2011SSRv..159..357Z}
{Zharkova} V~V, {Arzner} K, {Benz} A~O, {Browning} P, {Dauphin} C, {Emslie}
  A~G, {Fletcher} L, {Kontar} E~P, {Mann} G, {Onofri} M, {Petrosian} V,
  {Turkmani} R, {Vilmer} N and {Vlahos} L 2011 {\em \ssr\/} {\bf 159} 357--420
  (\textit{Preprint} \eprint{1110.2359})

\bibitem{2002SoPh..210....3L}
{Lin} R~P, {Dennis} B~R, {Hurford} G~J {\em et~al.\/} 2002 {\em \solphys\/}
  {\bf 210} 3--32

\bibitem{2014ApJ...780..176K}
{Kontar} E~P, {Bian} N~H, {Emslie} A~G and {Vilmer} N 2014 {\em \apj\/} {\bf
  780} 176 (\textit{Preprint} \eprint{1312.0266})

\bibitem{2012A&A...539A..43K}
{Kontar} E~P, {Ratcliffe} H and {Bian} N~H 2012 {\em \aap\/} {\bf 539} A43
  (\textit{Preprint} \eprint{1112.4448})

\bibitem{2011ApJ...731L..19F}
{Fleishman} G~D, {Kontar} E~P, {Nita} G~M and {Gary} D~E 2011 {\em \apjl\/}
  {\bf 731} L19 (\textit{Preprint} \eprint{1103.2705})

\bibitem{2011SSRv..159..301K}
{Kontar} E~P, {Brown} J~C, {Emslie} A~G, {Hajdas} W, {Holman} G~D, {Hurford}
  G~J, {Ka{\v s}parov{\'a}} J, {Mallik} P~C~V, {Massone} A~M, {McConnell} M~L,
  {Piana} M, {Prato} M, {Schmahl} E~J and {Suarez-Garcia} E 2011 {\em \ssr\/}
  {\bf 159} 301--355 (\textit{Preprint} \eprint{1110.1755})

\bibitem{2011SSRv..159..107H}
{Holman} G~D, {Aschwanden} M~J, {Aurass} H, {Battaglia} M, {Grigis} P~C,
  {Kontar} E~P, {Liu} W, {Saint-Hilaire} P and {Zharkova} V~V 2011 {\em \ssr\/}
  {\bf 159} 107--166 (\textit{Preprint} \eprint{1109.6496})

\bibitem{2013ApJ...767...83G}
{Graham} D~R, {Hannah} I~G, {Fletcher} L and {Milligan} R~O 2013 {\em \apj\/}
  {\bf 767} 83 (\textit{Preprint} \eprint{1302.2514})

\bibitem{2008ApJ...673..576X}
{Xu} Y, {Emslie} A~G and {Hurford} G~J 2008 {\em \apj\/} {\bf 673} 576--585

\bibitem{2011ApJ...730L..22K}
{Kontar} E~P, {Hannah} I~G and {Bian} N~H 2011 {\em \apjl\/} {\bf 730} L22
  (\textit{Preprint} \eprint{1102.3664})

\bibitem{2013ApJ...766...28G}
{Guo} J, {Emslie} A~G and {Piana} M 2013 {\em \apj\/} {\bf 766} 28
  (\textit{Preprint} \eprint{1303.1077})

\bibitem{1971SoPh...18..489B}
{Brown} J~C 1971 {\em \solphys\/} {\bf 18} 489--502

\bibitem{1972SvA....16..273S}
{Syrovatskii} S~I and {Shmeleva} O~P 1972 {\em \sovast\/} {\bf 16} 273

\bibitem{1978ApJ...224..241E}
{Emslie} A~G 1978 {\em \apj\/} {\bf 224} 241--246

\bibitem{2003ApJ...595L.115B}
{Brown} J~C, {Emslie} A~G and {Kontar} E~P 2003 {\em \apjl\/} {\bf 595}
  L115--L117

\bibitem{2014ApJ...787...86J}
{Jeffrey} N~L~S, {Kontar} E~P, {Bian} N~H and {Emslie} A~G 2014 {\em \apj\/}
  {\bf 787} 86 (\textit{Preprint} \eprint{1404.1962})

\bibitem{2003ApJ...595L.119E}
{Emslie} A~G 2003 {\em \apjl\/} {\bf 595} L119--L121

\bibitem{2015arXiv150503733K}
{Kontar} E~P, {Jeffrey} N~L~S, {Emslie} A~G and {Bian} N~H 2015 {\em ArXiv
  e-prints\/} (\textit{Preprint} \eprint{1505.03733})

\end{thebibliography}

\end{document}